\documentclass[final,3p,times,twocolumn,authoryear]{elsarticle}

\usepackage[cmex10]{amsmath}
\usepackage{array}
\usepackage{url}
\usepackage{amssymb}
\usepackage{color}
\usepackage{stfloats}
\usepackage{tabularx}
\usepackage[table]{xcolor}
\usepackage{booktabs}
\usepackage{multirow,graphicx}
\usepackage{tcolorbox}
\usepackage{colortbl}
\tcbuselibrary{skins}
\usepackage{caption}
\usepackage{amsmath}
\usepackage{natbib}

\journal{--}

\begin{document}

\begin{frontmatter}

\title{Mind the gap: quantification of incomplete ablation patterns after pulmonary vein isolation using minimum path search}

\author[label1]{Marta Nu\~nez-Garcia \corref{cor1}}
\ead{marta.nunez@upf.edu}
\cortext[cor1]{Corresponding author at: DTIC, Universitat Pompeu Fabra (office 55.121), Roc Boronat 138, E08018 Barcelona, Spain. Tel.: +34 93 542 1346}
\author[label1]{Oscar Camara}
\author[label2]{Mark D. O'Neill}
\author[label2]{Reza Razavi}
\author[label2]{Henry Chubb}
\author[label1]{Constantine Butakoff}

\address[label1]{Physense, Department of Information and Communication Technologies, Universitat Pompeu Fabra, Barcelona, Spain}
\address[label2]{Division of Imaging Sciences and Biomedical Engineering, King's College London, UK.}

% \tableofcontents

\begin{abstract}
Pulmonary vein isolation (PVI) is a common procedure for the treatment of atrial fibrillation (AF) since the initial trigger for AF frequently originates in the pulmonary veins.
A successful isolation produces a continuous lesion (scar) completely encircling the veins that stops activation waves from propagating to the atrial body. Unfortunately, the encircling lesion is often incomplete, becoming a combination of scar and gaps of healthy tissue. These gaps are potential causes of AF recurrence, which requires a redo of the isolation procedure.  
Late-gadolinium enhanced cardiac magnetic resonance (LGE-CMR) is a non-invasive method that may also be used to detect gaps, but it is currently a time-consuming process, prone to high inter-observer variability. In this paper, we present a method to semi-automatically identify and quantify ablation gaps. 
Gap quantification is performed through minimum path search in a graph where every node is a scar patch and the edges are the geodesic distances between patches.
We propose the Relative Gap Measure (\emph{RGM}) to estimate the percentage of gap around a vein, which is defined as the ratio of the overall gap length and the total length of the path that encircles the vein. Additionally, an advanced version of the \emph{RGM} has been developed to integrate gap quantification estimates from different scar segmentation techniques into a single figure-of-merit. Population-based statistical and regional analysis of gap distribution was performed using a standardised parcellation of the left atrium. We have evaluated our method on synthetic and clinical data from 50 AF patients who underwent PVI with radiofrequency ablation. 
The population-based analysis concluded that the left superior PV is more prone to lesion gaps while the left inferior PV tends to have less gaps ($p<0.05$ in both cases), in the processed data. This type of information can be very useful for the optimization and objective assessment of PVI interventions. 

\end{abstract}


\end{frontmatter}

\section{Introduction}
\label{sec:introduction}

Around 33.5 million people suffer from atrial fibrillation (AF) worldwide \citep{chugh2013worldwide}, the most frequent class of arrhythmia. 
The origin of AF is the appearance of rapid abnormal electrical signals activating the atrium in a disorganized way. Most of these abnormal electrical currents originate inside the pulmonary veins \citep{haissaguerre1998spontaneous}. 
Pulmonary vein isolation (PVI), which aims to electrically isolate the PVs from the main atrial body, is a common treatment, especially for patients not responding to medication. Radiofrequency ablation (RFA) is the most frequent technique for PVI, and involves the delivery of RF energy with an electrophysiological ablation catheter to create regions of scar tissue that will stop abnormal conduction.
Different ablation patterns are possible and their suitability is still under investigation \citep{bayer2016novel}. Nonetheless, PVI is considered the cornerstone of the procedure \citep{calkins20122012}, having been proven to terminate AF in many cases \citep{verma2005response,badger2010evaluation}. 
However, repeat procedures are frequently required, with the long-term success rate after catheter ablation ranging from 53.1\% with a single procedure to almost 80\% with multiple procedures (i.e. a redo) \citep{ganesan2013long}.

One of the most frequent reasons for this relatively low success rate is incomplete PVI, caused by the appearance of gaps in the created lesion.
In the case of RF ablation, lesion is formed at the tip of the catheter that is placed at discrete locations along the ablation path. 
Incomplete ablation lesions can appear due to that punctual (i.e. point by point) nature of the ablation, complexity of the atrial anatomy or complications during the procedure. The difficulty to access with a catheter may also lead to loss of contact and to non-transmural lesions \citep{glover2018preserved}. Additionally, atrial tissue can recover over time and therefore not all the acute lesions turn into chronic lesions.
These are potential factors for the presence of gaps after PVI \citep{kirchhof2016catheter}, increasing AF recurrence rates \citep{kuck2016impact}.

Gaps around PVs can be categorized into electrical (conduction) and anatomical (scar or lesion) ones. Conduction gaps are detected with intra-cavitary catheters during a redo procedure, and correspond to sites of electrical reconnection (i.e. high voltage in the electrocardiogram). The regional distribution of conduction gaps in the left atria after PVI was investigated by \cite{godin2013lessons} and \cite{galand2016localization}. 
On the other hand, lesion gaps are related to healthy tissue patches in the (ideally continuous) scar lesion. 
The relation between electrical and lesion gaps is still not fully understood; contradictory results can be found in the literature, from high \citep{bisbal2014cmr} to low \citep{spragg2012initial,harrison2015repeat} agreement in the location of both types of gaps. This disagreement is partially due to the lack of a consistent and objective way to detect and characterize ablation gaps, mainly relying on visual inspection of the data.

Ablation lesions (and therefore gaps) are typically identified using Late Gadolinium Enhancement Cardiac Magnetic Resonance (LGE-CMR). 
Gap detection in LGE-CMR remains challenging since there is a lack of consensus with respect to the most appropriate scar segmentation technique.
The interested reader is referred to \cite{pontecorboli2016use} for a recent review on scar detection in the atria with LGE-CMR.  
Most of scar segmentation techniques are semi-automatic and based on image thresholding: voxels are classified as scar or healthy tissue if an intensity-related value is above or below a given threshold, respectively. 
%Unfortunately, there is no consensus on how to estimate the most appropriate threshold. 
Many methods use a threshold calculated as a specific number of standard deviations (SD) above a reference value (e.g. mean intensity in the blood pool), but it is not clear which SD value is optimal: \cite{harrison2014cardiac} proposed 3.3 SD after histological validation, while the same research group did not find significant differences using 2, 3 or 4 SD in a previous study \citep{karim2013evaluation}. Recently, \cite{chubb2018reproducibility}, studying the reproducibility of LGE-CMR imaging, suggested that the intensity thresholding based on the mean and SD of the blood pool voxel intensity provided the most consistent scar identification between repeated LGE-CMR scans of the same patient.
Other threshold-based segmentation techniques are based on the Image Intensity Ratio (IIR), computed as the signal intensity divided by the mean of the blood pool. Different $IIR$ values have been proposed: while some authors \citep{dewire2014association,khurram2014magnetic} used a threshold of $IIR > 1.61$ to detect dense scar in pre-ablation cases, \cite{benito2016left} recently found an optimal $IIR > 1.32$ to define dense atrial fibrosis in post-ablation patients. A different strategy was taken by Bisbal et al. 2014, which adapted to atrial scar segmentation the standard thresholds currently used for left ventricle scar characterization in LGE-CMR (60 and 40 $\pm 5\%$ of the maximum intensity value to detect core zone, border zone and healthy tissue, respectively).

Once LA scar is detected, lesion gaps around PVs can be analysed. To the best of our knowledge, methods available in the literature are only based on visual inspection of scar segmentation results, providing qualitative and biased estimations of gap characteristics (i.e. number, position).
For example, \cite{peters2007detection} visually inspected LGE-CMR data to estimate the circumferential completeness and extent of ablation lesions after RF PVI, showing that in their dataset the completeness of the circumference was 88$\pm$11\% around the left inferior PV.
In \cite{badger2010evaluation} two independent and blinded observers visually identified and quantified scar and lesion gaps in LGE-CMR images concluding that complete circumferential PV scarring of all 4 PVs is difficult to achieve (only 7\% of patients  after the first procedure) but is related to better clinical outcome. Halbfass et al. \citep{halbfass2015lesion} took a similar approach (with three observers) to analyse lesion formation differences after PVI using two types of cryoballoon.  
The authors concluded that the number of patients with complete circular lesions did not differ significantly in the two groups. 

\begin{figure*}[t!]
\includegraphics[width=\textwidth]{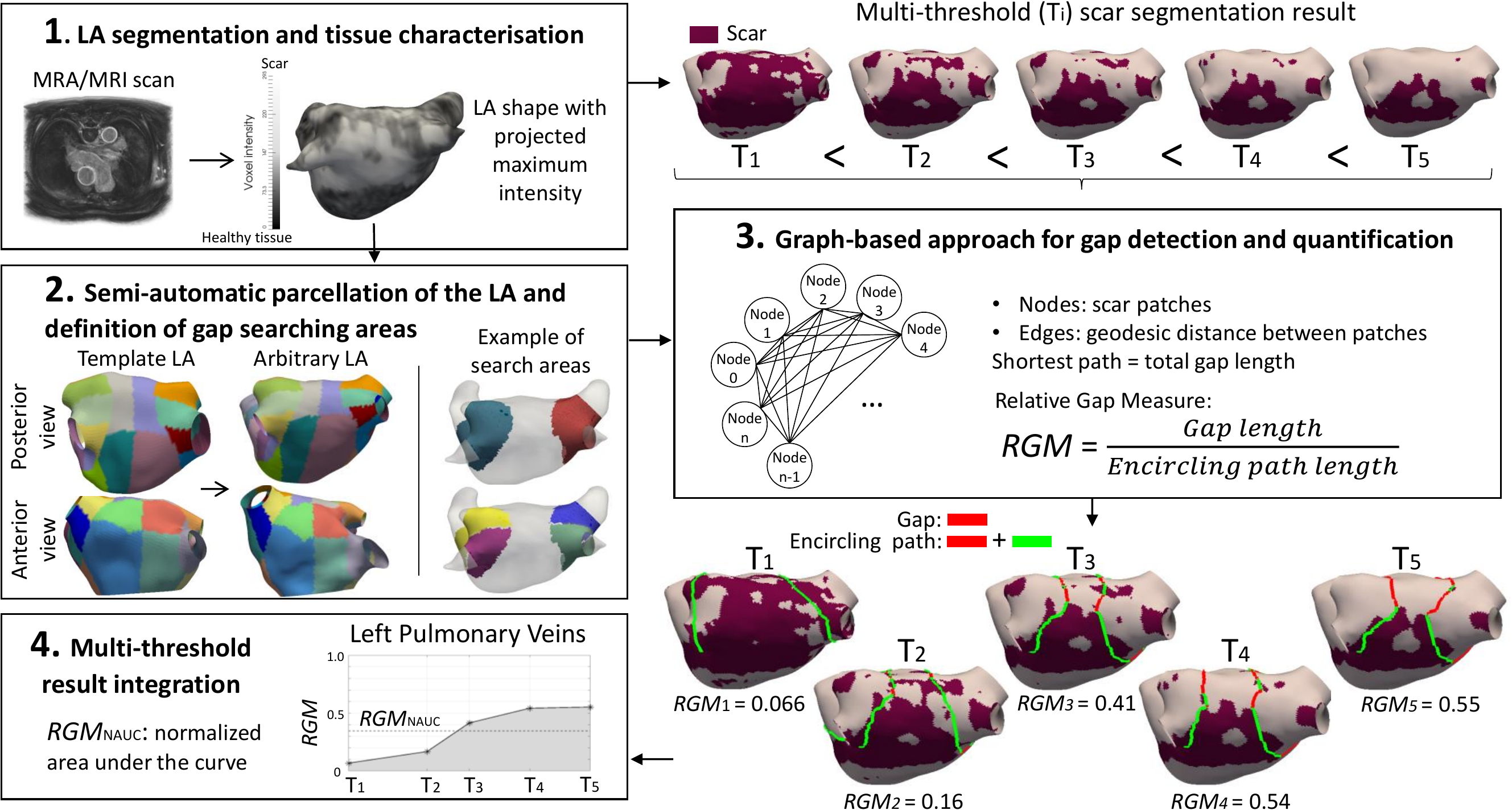} 
\caption{Pipeline of the proposed method. 1. Acquisition of left atrial (LA) anatomy and tissue information from Magnetic Resonance Angiongram (MRA) and Magnetic Resonance Imaging (MRI) scans respectively. A multi-threshold (T$_i$) scar segmentation approach is adopted, resulting in five different scar patterns. 2. Semi-automatic parcellation of the LA (each color represents a different region defined in our atlas) and definition of areas where the exploration of gaps will be performed. 3. Detection and quantification of gaps following a graph-based approach and using the \emph{Relative Gap Measure} (\emph{RGM}). 4. Combination of multi-threshold \emph{RGM} outcomes with the normalized area under the curve (RGM$_{NAUC}$). Shown is the RGM of only one search area (left PVs).}
    \label{fig:pipeline}
\end{figure*}

In this paper we present a consistent, objective and quantifiable definition of PVI gaps that is observer independent, once the scar has been segmented. 
We construct a graph where nodes are all scar patches around a given PV and edges are the distances between those patches (i.e. length of potential gaps). 
We estimate the shortest path in that graph and only the gaps belonging to that path are selected as true gaps. 
Then, the circumferential extent of the gap is measured as the fraction of the path's length that corresponds to gap, providing a Relative Gap Measure (\emph{RGM}). A preliminary version of the methodology was published in \citep{garcia2015quantification}, were gap quantification was performed using a 2D representation of the LA \citep{williams2017standardized}, so-called Standardised Unfold Map (SUM), which is obtained after flattening a 3D LA template with a pre-defined regional parcellation.  
Nevertheless, the obtained results were influenced by errors due to the metric distortion caused by the 3D-2D flattening. %This 2D approach was less accurate for gap quantification due to the distortion induced by the flattening, which could eventually miss tiny gaps. 
In this work, we propose an advanced version of our gap quantification method where all the steps have been considerably improved and with the following contributions: a more accurate distance calculation is provided since all measurements are performed directly on the 3D atrial mesh, allowing for an accurate visualization of the detected gaps on the atrial surface that could be used to guide a redo procedure; the development of an index representing the percentage of gap around a PV that can combine multiple segmentation results to compensate for the dependence of the process on scar segmentation accuracy; a detailed regional analysis of the gaps is additionally provided; and new PVI scenarios are considered as the typical ablation approach consisting in jointly isolating the two ipsilateral (i.e. same side) veins.
The method is highly reproducible only requiring minimal user interaction at the left atrial mesh processing step. To verify the proposed algorithm we generated a fully-controlled synthetic dataset of typical scar patterns around PVs. In addition, a clinical dataset composed of LGE-CMR studies of 50 AF patients was processed to study the regional distribution of gaps in this cohort.

\section{Methods}
\label{sec:methods}

The main steps of the algorithm are shown in Figure \ref{fig:pipeline} and explained in the following subsections.
Briefly, once a left atrial geometry has been segmented from imaging data, several scar segmentations are obtained by applying different thresholds ($T_{i}$) to image intensities (Step 1). Note that the thresholding in this step can be replaced by any scar segmentation algorithm.
In parallel, a semi-automatic parcellation of the LA is obtained to define gap search areas (Step 2). Subsequently, a graph is built for every scar segmentation result, where the nodes and edges correspond to scar patches and distances between them, respectively (Step 3). The graph is then analysed to quantify the amount of scar gap around each vein (or pair of veins), which is characterised by the \emph{Relative Gap Measure (RGM)}. Finally, \emph{RGM} values from the different $T_{i}$ threshold segmentations are combined into a single figure-of-merit ($RGM_{NAUC}$) (Step 4). 

The proposed method can be simplified if an accurate scar pattern is available: for each search area, only one graph would be processed resulting in a single \emph{RGM} value.

\subsection*{Patient data description}
\label{subsec:data}

Fifty patients undergoing first time ablation for atrial fibrillation between January 2014 and January 2016 at St Thomas' Hospital (London, UK) were included in this study. All of them provided written and informed consent and the study was approved by the National Research Ethics Service (South London Research Ethics Committee reference 08/H0802/68). 
For patients with a diagnosis of paroxysmal AF and in sinus rhythm, a point-by-point wide area circumferential ablation (WACA) achieving PVI was performed using 8Fr irrigated SmartTouch catheter (Biosense Webster), or 8Fr irrigated TactiCath catheter (St Jude). Target ablation parameters were $>$5g for at least 15 seconds per RF delivery location. Power was 30W throughout except on the posterior wall, where it was limited to 25W. 
For patients presenting with persistent AF, a WACA was performed followed by additional ablation lesion sets (mitral line, roof line, inferior posterior line, complex fractionated electrogram ablation) as a step-wise ablation. 

CMR imaging was performed on a 1.5 T MR-scanner (Ingenia, Philips Healthcare, Best, Netherlands) three months after the ablation. 
3D inversion recovery spoiled gradient echo acquisition was performed with coverage to include the whole LA in axial orientation, using the following MR acquistion parameters: TR 5.5 ms, TE 3.0 ms, flip angle 25$^{\circ}$, low-high k-space ordering, respiratory and ECG-triggering (end atrial diastole, maximum 120 ms acquisition window), 1.3 x 1.3 x 4 $mm^{3}$ (typically 50 slices per acquisition), SPIR fat suppression.
Gadolinium-Based Contrast Agents (GBCA) dose was 0.2 ml/kg Gadovist (Bayer HealthCare Pharmaceuticals, Berlin, Germany). Acquisitions were commenced 30 min post gadolinium-based contrast agent administration \citep{chubb2018optimization}.
A gated magnetic resonance angiogram (GMRA) 3D dataset was also acquired as a high contrast template. The acquisition was commenced 90 seconds after the start of a slow infusion of GBCA at 0.3ml/second \citep{groarke2014feasibility} with the same coverage and parameters (previously mentioned) as the subsequent LGE acquisitions.

\subsection{Left atrial segmentation and tissue characterisation}
\label{subsec:scar_segment}

Left atrial geometries were obtained from the GMRA images with a standard region-growing segmentation technique initialized by manually placed seeds, available in the Open-Source ITK-SNAP\footnote{www.itksnap.org} software \citep{yushkevich2006user}.
The resulting LA segmentation was then rigidly registered (3 translations and 3 rotations, \citep{denton1999comparison}) to the LGE-CMR images and a triangular mesh was built using the classical marching cubes algorithm. Voxel intensities from LGE-CMR images were mapped onto the obtained LA surface mesh using the maximum intensity projection (MIP) technique: images were sampled along the normals on both sides of the surface mesh, assigning the maximum intensity value to the corresponding vertex on the LA mesh. The depth of the sampling was set to 3 mm (0.2 mm between samples). This value is considered to be large enough to overcome potential segmentation errors while at the same time sufficiently small to avoid going too far from the atrial wall \citep{knowles20103,harrison2015advances}. 

Scar tissue was extracted from LGE-CMR images using intensity thresholding based on the SD of the blood pool intensity since it was the approach that demonstrated better reproducibility \citep{chubb2018reproducibility}.
In order to reduce the influence of the SD choice on gap quantification, we identified scar with five different thresholds: 2, 3.3, 4, 5, and 6 SD above the mean intensity of the blood pool. These thresholds were chosen based on the work of \cite{harrison2014cardiac}, where the authors obtained the best segmentation accuracy with the 3.3 SD value. According to their results, a non-linear behavior between scar patterns and the choice of SD parameter was observed: SD values lower than the optimal one overestimated the scar in a higher proportion than higher values underestimated it. For this reason, three different SDs were chosen above 3.3 SD and just one below it. 
In consequence, for each subject we obtained five different scar patterns that were separately processed in the gap detection step.  
Figure \ref{fig:pipeline} (top row) shows the scar patterns obtained by different thresholds in an exemplar case, where scar patches split or disappear (i.e. new gaps appear or merge, respectively) with larger thresholds ($T_{1}<T_{2}<T_{3}<T_{4}<T_{5}$).

\subsection{Semi-automatic parcellation of the LA}
\label{subsec:parcellation}

A standardised parcellation of the segmented LA surface is necessary to define the areas where gaps will be searched and to have a common reference system for population-based analysis. We used a parcellation created on a template left atrium (atlas mesh), as we proposed in \cite{williams2017standardized}. 
An extra separation between left and right parts of the LA was added, leading to 28 regions. This was required to test additional ablation patterns commonly used in clinical routine (see Step 2 in Figure \ref{fig:pipeline}).

\begin{figure*}[t]
\includegraphics[width=0.95\textwidth]{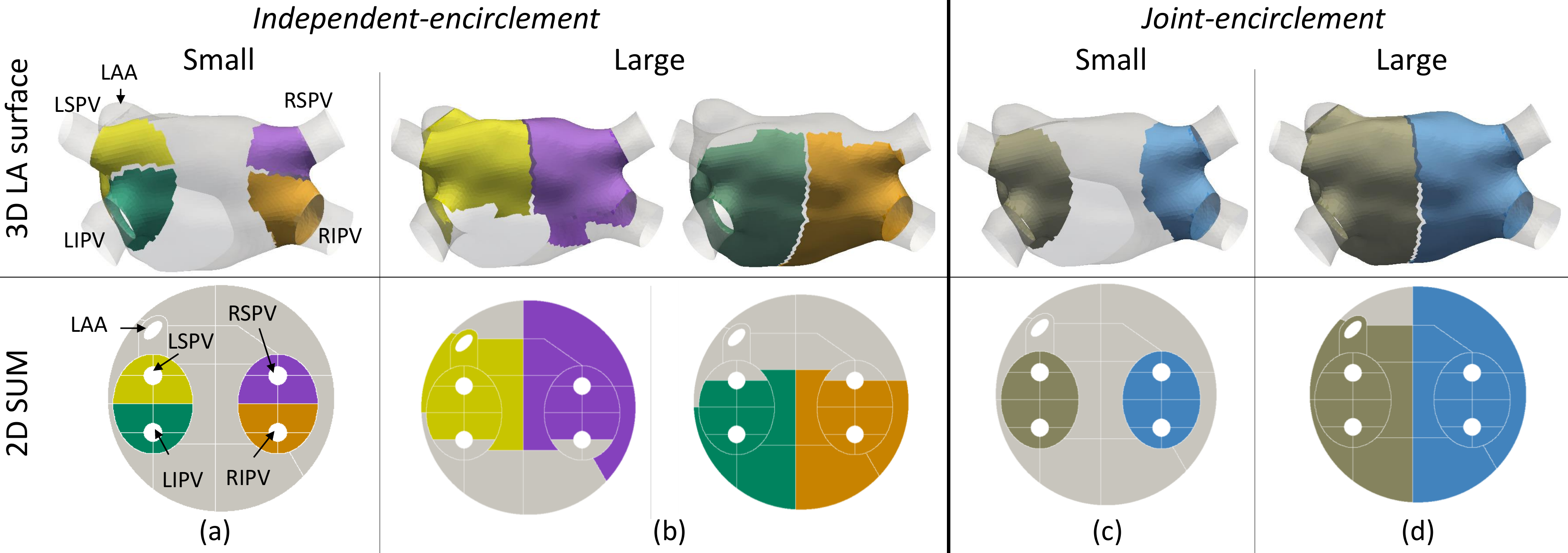} 
    \caption{Different gap search areas depending on the pulmonary vein isolation strategy, illustrated by different colors on 3D LA surface (top) and 2D SUM (bottom) representations. From left to right: \emph{independent-encirclement} with small (a) and large (b) regions; \emph{joint-encirclement} with small (c) and large (d) regions.}
    \label{fig:areas_of_influence}
\end{figure*}

\subsubsection*{Surface mesh pre-processing}
To transfer the defined parcellation in the atlas mesh to any arbitrary LA surface mesh several pre-processing steps were required. The first step aimed at standardising the LA shape by only keeping its main cavity after semi-automatically cutting connected sub-structures such as the PVs, the left atrial appendage (LAA) and the mitral valve. This cutting process only requires to manually place 5 seeds near the ending points of the PVs and the LAA. The reader is referred to \cite{tobon2015benchmark} for more details on this method. 
Once atrial meshes were standardised they were registered to the atlas mesh. The registration stage was composed of an affine transformation followed by a non-rigid registration based on currents \citep{durrleman2014morphometry}. Region labels of the atlas were then transferred to the co-registered segmented LA mesh with a closest-point mapping. Afterwards, region labels were also brought to the original (non-registered) LA mesh, where the subsequent analysis was performed. Additionally, we used the 2D SUM representation \citep{williams2017standardized} of the LA template for visualization purposes.

\subsubsection*{Gap search areas}

\begin{figure*}[ht]
\centering
\includegraphics[width=0.90\textwidth]{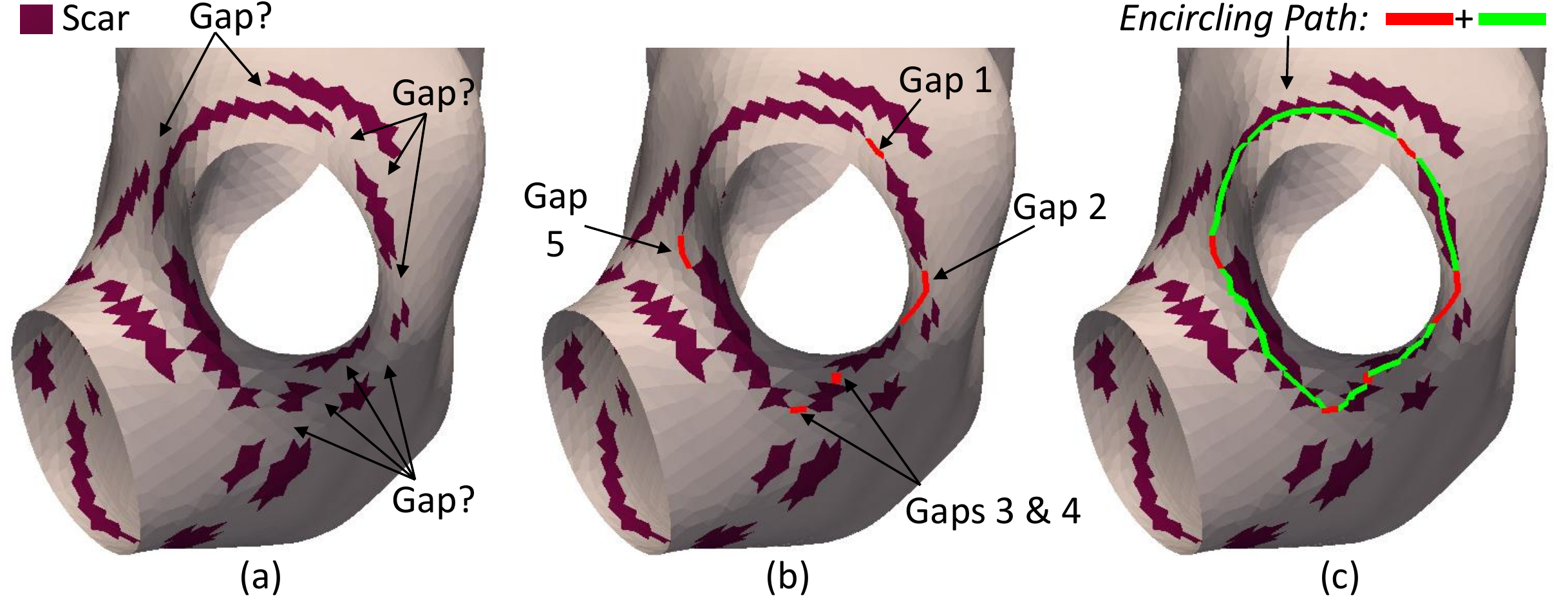} 
    \caption{(a) Example of synthetic data with a patchy scar that suggests many potential gaps. (b) According to our definition, not all of them are true gaps, only the ones shown in red. (c) The complete \emph{Encircling Path} is the concatenation of gaps and paths connecting the gaps across scar patches (red and green paths, respectively).}
    \label{fig:potential_gaps}
\end{figure*}

\begin{figure*}[ht]
\begin{center}
\includegraphics[width=\textwidth]{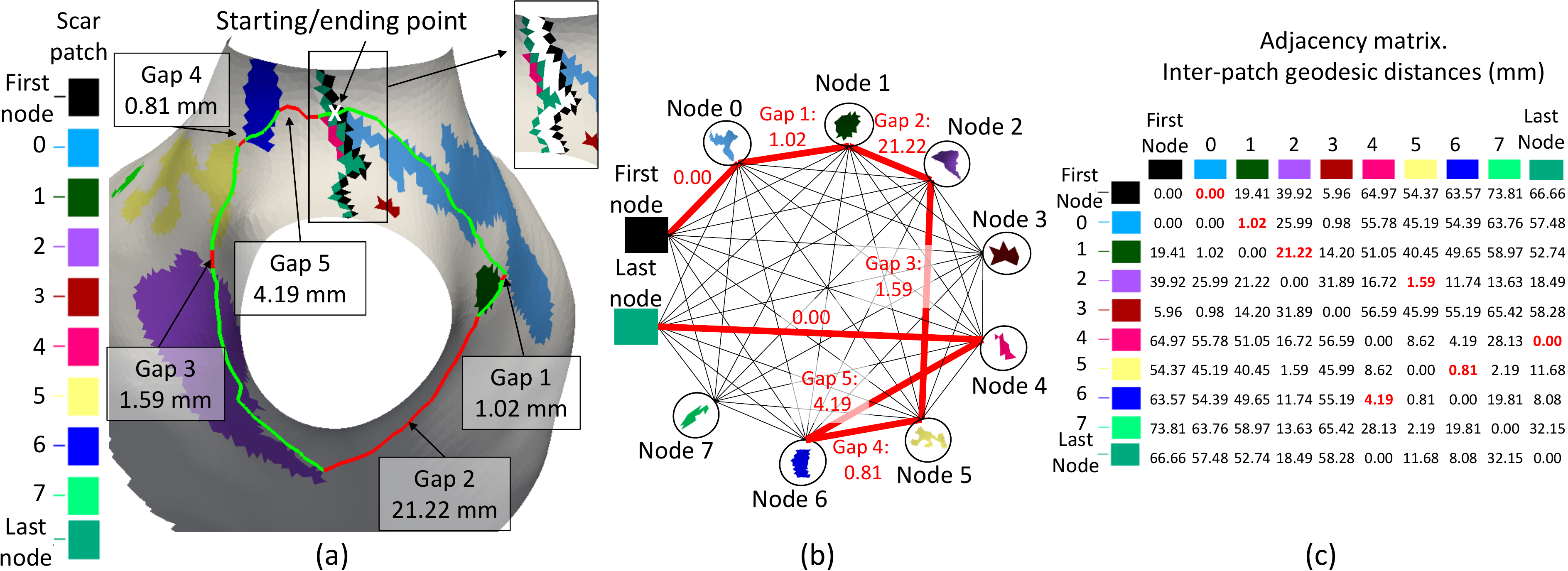} 
\end{center}

	\caption{Example of gap detection. (a) Detail of the search area around a PV showing all scar patches in different colors. The artificial scar patches (green and black) created along the cut are also shown, as well as the detected gaps (red) and the rest of the \emph{Encircling Path} (green). The constructed graph (b) and the adjacency matrix (c) are used to determine the shortest path around the PV (colored in red) going from the starting to the ending point. Distances are reported in mm.} 
   \label{fig:graphs_example}
\end{figure*}

\begin{figure*}[ht!]
\includegraphics[width=0.99\textwidth]{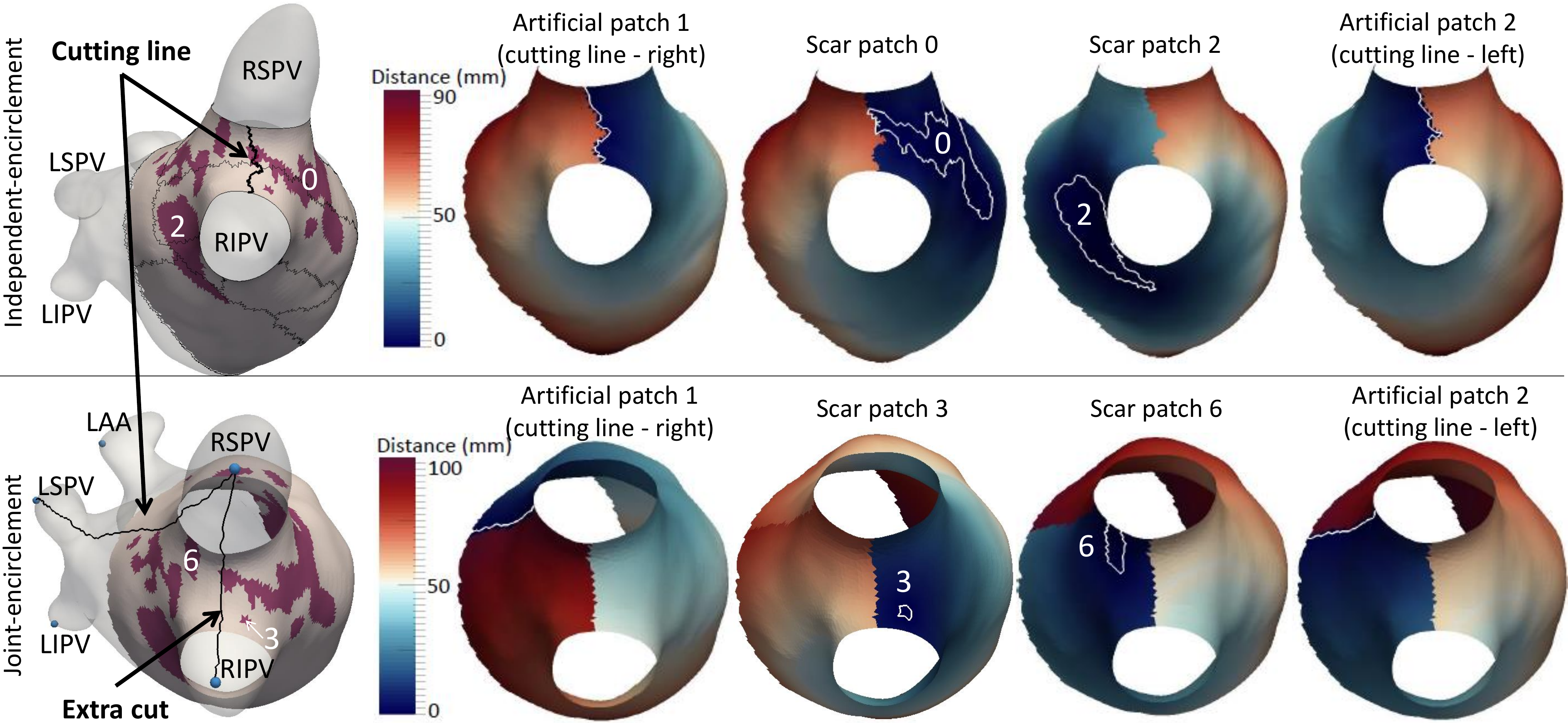} \\
 \caption{Several distance maps corresponding to an exemplar case (leftmost column). 
The first row shows the \emph{independent-encirclement} approach and the second row the \emph{joint-encirclement} approach. Each column displays a distance map corresponding to a different scar patch (numbers are patch identifiers) whose contour is also shown in white. 
The leftmost and rightmost distance maps correspond to the distance transforms of the cutting line (i.e. artificial scar patches). Note the effect of the extra cut needed in the \emph{joint-encirclement} approach to avoid that the \emph{Encircling Path} surrounds only one PV. LSPV = Left Superior PV; LIPV = Left Inferior PV; RIPV = Right Inferior PV; RSPV = Right Superior PV; LAA = Left Atrial Appendage.
}   
\label{fig:dts}
\end{figure*}

The next step of the pipeline defines the areas where gaps will be searched. These areas are related to the chosen PVI strategy. In this work we considered the two most common PVI scenarios in clinical routine:

\begin{itemize}
\item \emph{Independent-encirclement}: The four PVs are independently isolated by creating PV-specific continuous lesions that completely surround each of them;
\item \emph{Joint-encirclement}: the two ipsilateral veins (i.e. on the same side, right or left PVs) are jointly isolated by a lesion that simultaneously encircles the two of them. 

\end{itemize}

The two PVI scenarios require a different definition of the gap search area, as illustrated in Figure \ref{fig:areas_of_influence}. 
Additionally, different sizes of the search areas were considered:
small regions near to the veins or larger regions that would give a more global view. 
Since oftentimes ablation lines tend to deviate far from PV ostia (e.g. due to undesired catheter movement), we chose to perform the subsequent gap analysis only in the largest search areas (Figure \ref{fig:areas_of_influence} (b) and (d)).

\subsection{Graph-based gap identification and quantification}

The developed method is based on our definition of a potential gap as the shortest geodesic distance (i.e. length of the shortest curve between two points on a mesh, such that the curve lies on the surface \citep{mitchell1987discrete}) between two disjoint scar patches within a search area. Many paths exist that surround a PV; we define the \emph{Encircling Path} as the closed path that encircles a PV (or two same-side PV together) with the minimum amount of gap, as can be seen in Figure \ref{fig:potential_gaps}. In that way, only potential gaps (i.e healthy tissue between scar patches) belonging to the \emph{Encircling Path} are classified as true gaps. Figure \ref{fig:potential_gaps} shows how the proposed method is able to consistently identify the true gaps in a situation where the visual detection and quantification of gaps would be challenging.

We defined a quantitative index to represent the amount of gap in each of the search areas, so-called \emph{Relative Gap Measure (RGM)}, which is obtained as follows:

\begin{equation*}
RGM = \frac{\text{\emph{Gap length}}}{\text{\emph{Encircling Path length}}},
\end{equation*} 

\noindent where $Gap~length$ (GL) adds up all gap lengths along the \emph{Encircling Path}; and the $Encircling~Path~length$ corresponds to the length of the complete closed loop.
%also {\color{blue} includes the rest (non-gap) of the path, 
%as shown in Figure \ref{fig:potential_gaps}. 
The $RGM$ ranges from 0 to 1 indicating that the vein is completely encircled if $RGM=0$ and that there is no scar around the vein if $RGM = 1$.
% includes geodesic distances {\color{blue} between consecutive gaps in the path, as shown in Fig. \ref{fig:potential_gaps}.} %across scar patches. 

To find the \emph{Encircling Path} and compute the \emph{RGM}, a graph was built (see Figure \ref{fig:graphs_example} (b)) %we constructed a graph 
where each node was a scar patch and edges were the minimum geodesic distances between pairs of scar patches (i.e. minimum inter-patch distance), that is, the length of the potential gaps. The graph was complete, in the sense that every pair of nodes were connected. The edges of the graph were estimated using the distance transform \citep{danielsson1980euclidean,fabbri20082d} that is computed considering each scar patch separately: it assigns to every point in the mesh a value indicating its distance to the nearest point in the scar patch under study %Note that if the point belongs to the scar patch the distance assigned is 0 
(see Figure \ref{fig:dts}).
The two points corresponding to the extremes of the selected path (i.e. associated to the minimum inter patch distance) were identified as gap limits. 

Using the graph and its associated adjacency matrix (Figure \ref{fig:graphs_example} (b) and (c) respectively) we can determine the \emph{Encircling Path}, but with the following considerations: %need to be taken into account: 
the starting and ending points in the path must coincide to obtain a closed path; and an orientation must be imposed in the calculation of the path to guarantee that the path completely surrounds the vein.
We fulfilled these requirements opening the mesh, by introducing a cutting line coinciding with the atrial region boundaries defined on the LA template. 
This cutting line was implemented by disconnecting its cells after adding duplicate points: points with the same coordinates but different connectivity. 
%As the points in the cutting line are disconnected, the \emph{Encircling Path} can not cross that limit and it constitutes in fact the extremes of the path.
The \emph{Encircling Path} cannot cross the cutting line since duplicate points are disconnected, becoming the extremes of the path.
We then introduced two artificial scar patches (and corresponding nodes in the graph) at the cutting line (one to each side) defining the first and the last nodes in our graph. These nodes are the only ones not directly connected in the built graph (see Figure \ref{fig:graphs_example} (a) and (b)). The detection of the \emph{Encircling Path} is then based on finding the minimum distance between the two extreme nodes using the corresponding adjacency matrix (Figure \ref{fig:graphs_example} (c)).  

\begin{figure*}[th]
\includegraphics[width=\textwidth]{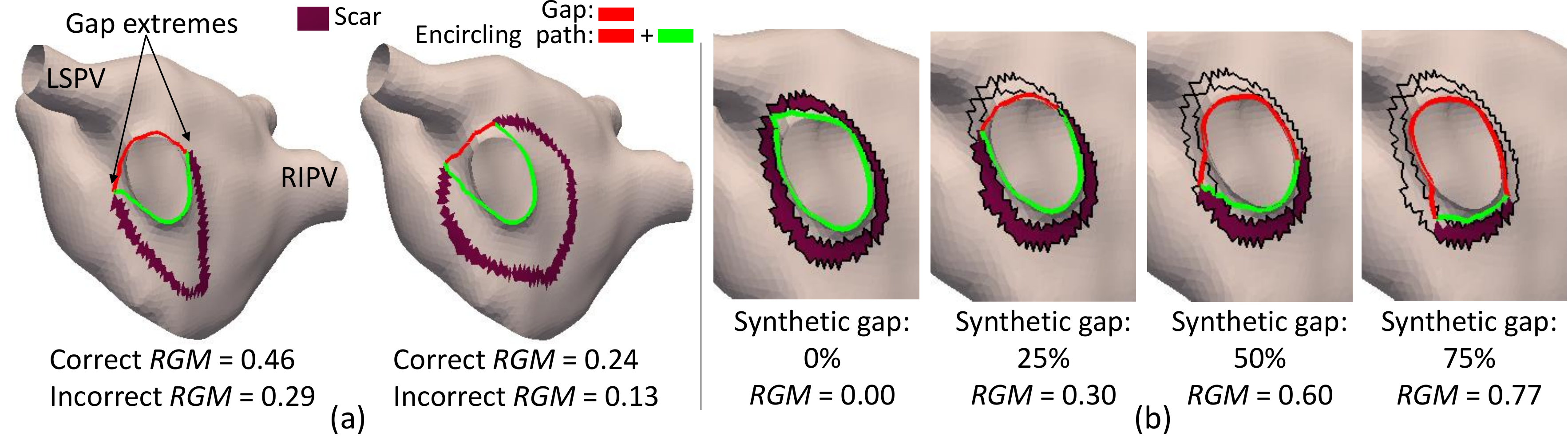} 
\caption{Examples of \emph{Relative Gap Measure} (\emph{RGM}) estimations on synthetic data where scar was added around the left inferior PV. In each example the scar pattern is shown, together with the \emph{Encircling Path}, which has two parts: gaps (red) and non-gap paths (green).
%scar-related path (green). 
(a) Examples with scar patterns distant from the PV ostia, illustrating why it is appropriate to compute the \emph{Encircling path} as a concatenation of gaps and geodesics connecting the gaps, i.e. using gap extremes
%the non-gap length}
%scar length 
%using gap extremes 
 (correct \emph{RGM} result), independently of scar shape (incorrect \emph{RGM} result). (b) Examples of scar patterns with increasing \emph{RGM}. The black contours correspond to an initial scar patch fully encircling the PV at a certain distance from the extreme of the PV. Then, in each example, a percentage of this patch is marked as gap (from left to right: 0\%, 25\%, 50\%, and 75\%, respectively).} 
    \label{fig:syn}
\end{figure*}

The first row in Figure \ref{fig:dts} displays several distance maps where the effect of the cut is shown as a sudden color change in the cutting line.
%noticeable (sudden color change in the cutting line). 
In the \emph{joint-encirclement} approach (second row in Figure \ref{fig:dts}) an additional cut along the line connecting the two involved veins had to be added to prevent the \emph{Encircling Path} surrounding only one vein. %, i.e. the \emph{Encircling Path} should not traverse that area. 
This cut was computed using the seeds manually placed in the veins during the surface mesh pre-processing step. %(see the second row in Figure \ref{fig:dts}). 

The distances between artificial and real scar patches needed a particular approach to avoid a non-closed \emph{Encircling Path}. Naturally, all points of the cutting line are candidates as starting/ending points of the \emph{Encircling Path}. We built a graph for each possible pair of starting/ending points and selected the %optimal option (i.e. with the smallest gap length).  
ones with the smallest gap length.
The shortest path between all possible starting/ending point candidates, was obtained applying the Dijkstra algorithm \citep{dijkstra1959note}. 
As a direct output of the algorithm we also obtained the number of gaps (and their lengths) in each \emph{Encircling Path}.

Estimation of the \emph{RGM} index also requires computing the length of the non-gap section in the \emph{Encircling Path}. Scar tissue can present any type of morphology around each PV, but in order to compute the percentage of encirclement completeness, scar morphology is not relevant, and only the corresponding proportion of the PV boundary that is isolated matters.
%Computing distances across scar patches %(intra scar length) 
%was not obvious due to the heterogeneity of scar characteristics around PVs. 
One possibility would be to estimate the geodesics following the scar shape, but this approach would underestimate the \emph{RGM}. Figure \ref{fig:syn} (a) shows a synthetic example where the gap covers about half of the PV and the scar is quite distant from the PV. %If scar length would be estimated following its shape, 
If the non-gap part of the \emph{Encircling path} would be estimated following the scar shape, as its length is quite large,
the obtained \emph{RGM} would be 0.29, which does not correspond to the correct proportion of gap around the vein. We solved this issue by computing the %scar length 
non-gap part of the \emph{Encircling path}
as the geodesics
%geodesic distance 
between the two points in the same scar patch identified as gap extremes. The \emph{Encircling path} is then a concatenation of gaps and geodesics connecting consecutive gaps.
In the same example, with our approach the \emph{RGM} is 0.46, which is a more reasonable value according to our perception.    
Similarly, on the right example in Figure \ref{fig:syn} (a), the \emph{RGM} considering the real shape of the scar would be equal to 0.13 suggesting again a smaller gap.

\begin{figure*}[h!]
\includegraphics[width=\textwidth]{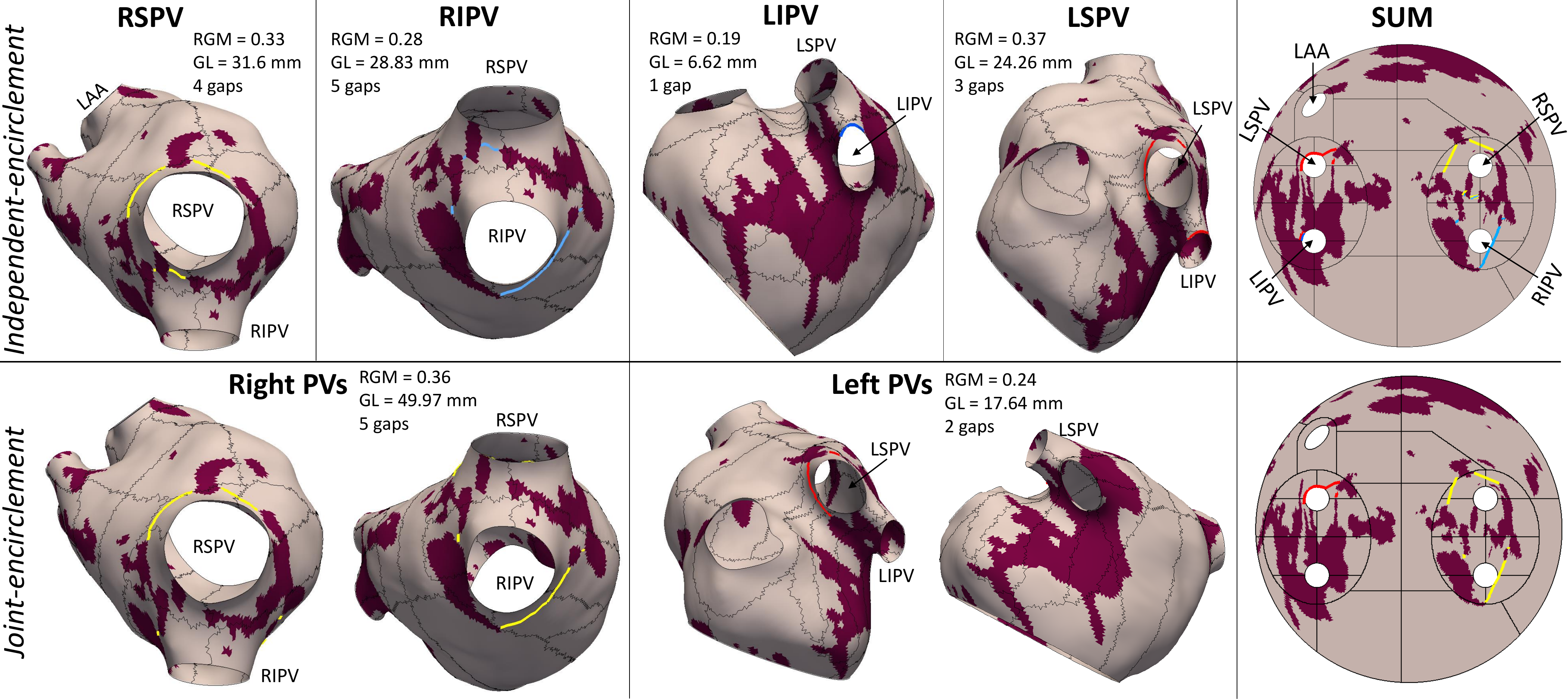} 
    \caption{Example of gap quantification in clinical data with the two studied pulmonary vein isolation strategies: \emph{independent-} and \emph{joint-encirclement} in the first and second rows, respectively. Several views of the atria are shown as well as the corresponding SUM (right). The gaps detected in the different search areas (i.e. RSPV, RIPV, LIPV, LSPV, right PVs and left PVs) are shown with different colors. Notice that some of the gaps are part of the encircling path of the two veins in the same side. In that situation, two parallel gap paths are shown but there is only one gap path (the shortest one) which is the same for both veins.}
    \label{fig:individual_analysis}
\end{figure*}

Besides estimating the \emph{RGM}, the graph-based gap step also provides additional measures to quantify gaps for each PV and pair of veins such as the (accumulated and regional) number and length of gaps. Figure \ref{fig:individual_analysis} shows these gap quantification measures in an exemplar case.

\subsection{Multi-threshold result integration}

As mentioned in Section \ref{subsec:scar_segment}, we used a multi-threshold protocol for scar segmentation, generating five different scar patterns per case. The proposed gap quantification pipeline was applied to each segmented scar separately and the results were then integrated as follows.
 
\begin{figure*}[h!]
\includegraphics[width=0.99\textwidth]{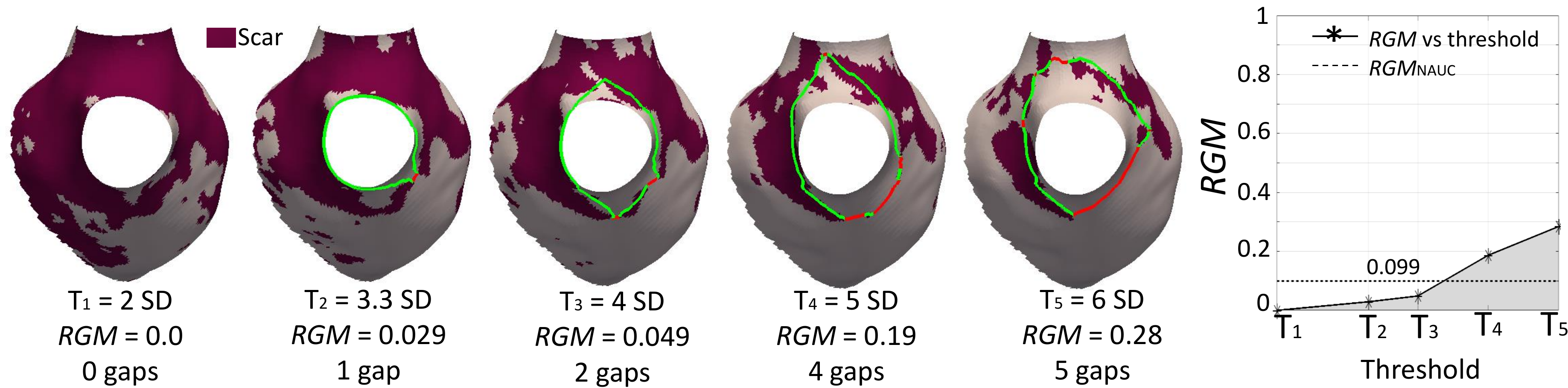} 
    \caption{Gap quantification result in a given search area combining multiple scar segmentations. From left to right, an increasing threshold is used which results in an increasing amount of gap and \emph{RGM}. The \emph{Encircling Paths}, composed by gap (red) and non-gap (green) sections, are shown for every scar segmentation.
    %: gaps (red) plus {\color{blue} non-gap paths (green).} %scar-related path (green). 
    The rightmost figure shows the \emph{RGM} vs threshold curve as well as the $RGM_{NAUC}$ value.} 
    \label{fig:evol}
\end{figure*}

\emph{RGM} measurements were combined computing the normalised area under the curve (NAUC) of the plot relating the \emph{RGM} and threshold values (let us denote this measure by $RGM_{NAUC}$), as can be seen in Figure \ref{fig:evol}. The AUC of the \emph{RGM}-derived curve was estimated using the trapezoidal rule that was then normalized by the maximum area; given that $RGM \epsilon [0,1]$, it was simply the threshold's range ($T_{5}-T_{1}$). Let $f: T \epsilon [T_1, T_5] \rightarrow [0,1]$ be the function defined on the interval of thresholds assigning the \emph{RGM} value to each threshold. Then, $RGM_{NAUC}$ is obtained as follows:

\begin{equation*}
\begin{split}
RGM_{AUC} = \int_{T_1}^{T_5}f(T) dT \approx \\ 
\approx \sum_{k=1}^{4} \frac{f(T_{k+1})+f(T_{k})}{2} (T_{k+1} - T_{k}),
\end{split}
\end{equation*} 
and
\begin{equation*}
RGM_{NAUC} = \frac{RGM_{AUC}}{T_{5}-T_{1}}
\end{equation*}

Besides the $RGM_{NAUC}$ index, the multi-segmentation approach provides statistics (mean and standard deviation) on gap quantification measurements (number and length of gaps) across the different segmentations in each search area.

\section{Experiments and Results}
\label{sec:exper}

\subsection{Synthetic data}
%A verification of the proposed algorithm was performed on synthetically generated data consisting in varying scar patterns on LA template mesh. 
The performance of the proposed gap detection and quantification methodology was verified on synthetically generated data with varying scar patterns on the LA template mesh. Figure \ref{fig:syn} (b) shows several synthetic scar patterns where a scar fully encircling the PV was synthetically generated at a distance between 2 and 4 mm from the PV ostium (this region is marked as a black contour in the figure). %Out of this full scar, we 
We then removed 0\%, 25\%, 50\%, 75\% and 100\% of the synthetic scar, consequently generating gaps of different size. %path as such, generating in that way gaps of different size. 
The results of our gap quantification method were \emph{RGM} = 0.00, 0.30, 0.60, 0.77 and 1.00, respectively (the last case is not shown in the figure).
%It can be observed how the result slightly differs from the expected outcome according to the percentage of the full scar included. 
\emph{RGM} estimations slightly differ from the percentage of scar removed (maximum deviation of 10\%). 
This is explained by our definition of gap as the smallest portion of healthy tissue around the PV (therefore not following the black contour) and due to the computation of the non-gap section of the \emph{Encircling path} using gap extremes (also not following the black contour which corresponds to the scar patch in this case).

\subsection{Clinical data}

We applied our method to an initial set of 50 patient LA from which 5 cases were excluded due to morphological incompatibilities (number of PVs different than 4) with our method. 

Figure \ref{fig:individual_analysis} shows gap quantification results obtained on a clinical dataset. 
It can be observed that gap detection depends on the chosen ablation strategy. For instance, gaps in the \emph{independent-encirclement} approach (in the right carina and around the LIPV) are not gaps in the \emph{joint-encirclement} approach. %It can be observed how the SUM highly
The 2D SUM representation considerably facilitates the overall visualization of the results.

The use of a standard LA division allowed population studies once the gap quantification pipeline was applied to our entire clinical database. The amount and regional distribution of gaps in the whole population were analysed in two phases: firstly, we studied the $RGM_{NAUC}$ and the quantity of gaps in each search area; and secondly, the specific position and length of gaps in the 28 standard regions in our atlas was inspected.
We additionally evaluated the differences in gap quantification obtained with single- vs multi-threshold approaches for scar segmentation.

\begin{table}[t]
     \caption{Mean and standard deviation of the $RGM_{NAUC}$, the gap length (GL), and number ($\#$) of gaps detected in our dataset. Statistically significant differences (in bold) were found in the LIPV and LSPV. RSPV = Right Superior PV; RIPV = Right Inferior PV; LIPV = Left Inferior PV; LSPV = Left Superior PV.}
\resizebox{\columnwidth}{!}{% 
\begin{tabular}{c|cccc|cc}
& \multicolumn{4}{c|}{\textbf{\emph{Independent-encirclement}}} & \multicolumn{2}{c}{ \textbf{\emph{Joint-encirclement}}} \\ 
    &  \textbf{RSPV} & \textbf{RIPV}& \textbf{LIPV}& \textbf{LSPV} & \textbf{Right} & \textbf{Left}\\
    \hline
    \textbf{Mean $\mathbf{RGM_{NAUC}}$} & 0.303 & 0.331 & \textbf{0.208} & \textbf{0.384} &  0.261 & 0.249\\
    \textbf{SD $\mathbf{RGM_{NAUC}}$} & 0.269 & 0.317 & 0.207 & 0.271  & 0.224 & 0.199\\
    \hline
   \textbf{Mean GL (mm)}& 17.69 & 14.85 & 8.66 & 16.37  & 27.74 & 21.86 \\
   \textbf{SD GL (mm)}& 14.72 & 12.97 & 7.42 & 10.76 & 23.22 & 16.01 \\
    \hline
    \textbf{Mean \# gaps} & 1.56 & 1.36 & \textbf{1.16} & \textbf{1.73} & 2.34 & 2.00\\
    \textbf{SD \# gaps} & 0.85 & 0.89 & 0.64 & 0.81 & 1.63 & 1.20\\
    \hline
\end{tabular}%
}
\label{tab:means_together}
\end{table}

\begin{figure}[t]
\includegraphics[width=0.99\columnwidth]{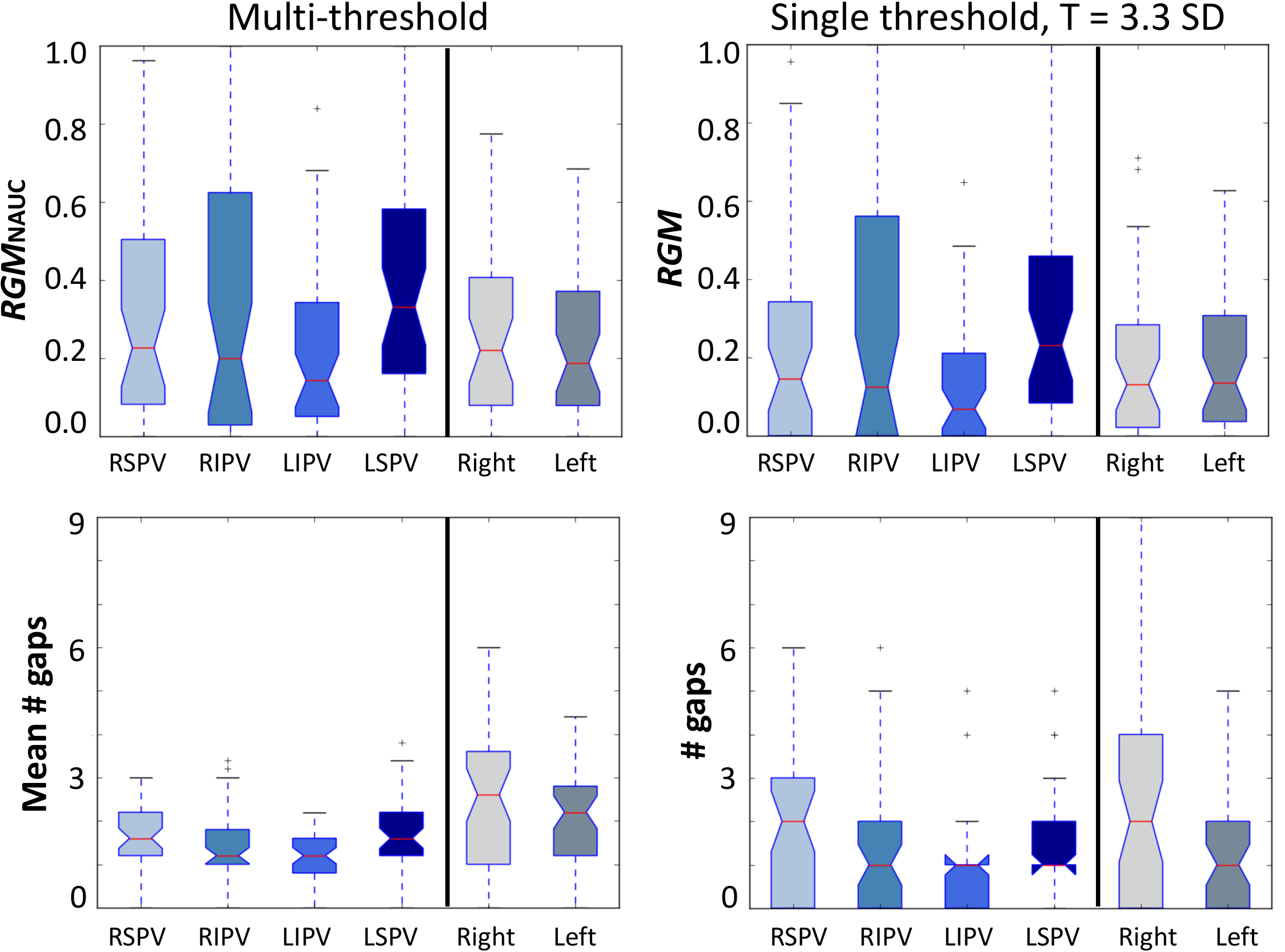}
\caption{Left: gap quantification provided by the proposed multi-threshold approach: distributions of the $RGM_{NAUC}$ (top) and the average number of gaps (bottom) around the different veins. On the right, results corresponding to a single threshold approach (3.3 SD): distributions of the \emph{RGM} (top) and the number of gaps (bottom).}
\label{fig:boxplots_mean}
\end{figure}

\begin{figure*}[t]
\includegraphics[width=\textwidth]{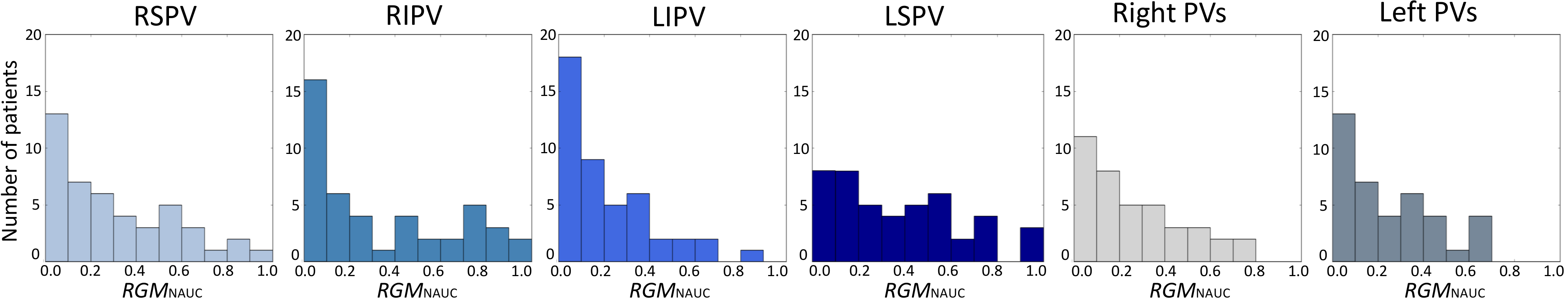}
\caption{Histograms of the $RGM_{NAUC}$ (normalized area under the curve of regional gap measurement for multiple-threshold segmentations) in all search areas, representing the amount of patients within a given $RGM_{NAUC}$ interval. Ideal ablation should locate most patients in the first bin (0.0 to 0.1 $RGM_{NAUC}$), in a skewed right shape.}
\label{fig:hists}
\end{figure*}

\begin{figure}[t]
   \includegraphics[width=\columnwidth]{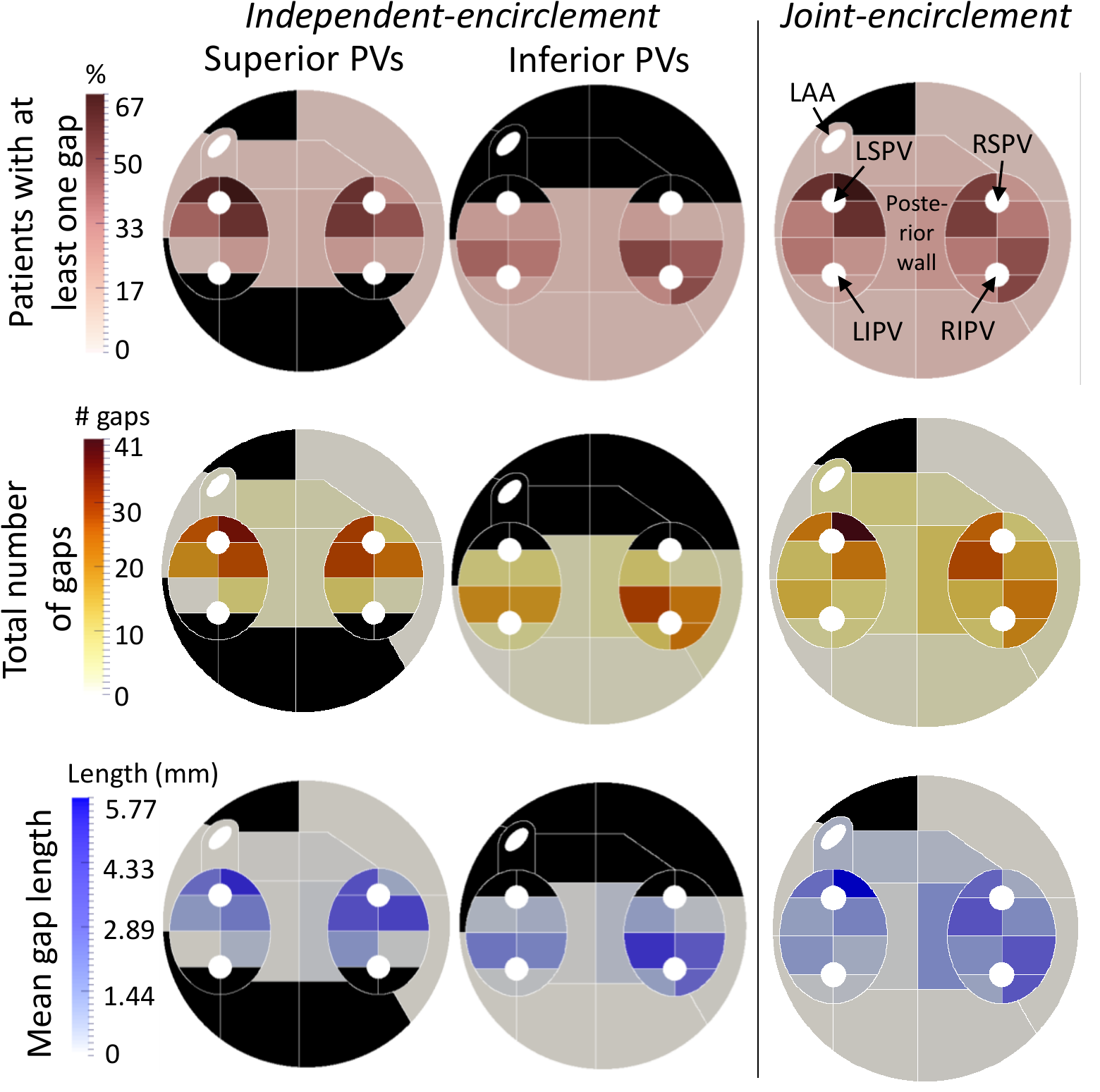}
   \caption{Results from the regional patient population study: for each region, the top, middle and bottom row show the percentage of patients with at least one gap, the total number of gaps, and the mean gap length, respectively. The first two columns correspond to the \emph{independent-encirclement} approach (superior and inferior veins are shown separately because of the overlapping regions) and the third column corresponds to the \emph{joint-encirclement} approach. Black regions represent the ones excluded in each approach.}
    \label{fig:SUM_percentages_colors}
\end{figure}

Table \ref{tab:means_together} shows, for each (or pair of) PV, statistics (mean and standard deviation) of the $RGM_{NAUC}$, the average (of multiple thresholds) gap length and number of gaps. 
Similarly, the left column in Figure \ref{fig:boxplots_mean} shows the boxplot corresponding to the $RGM_{NAUC}$ and to the average number of gaps (top and bottom in the figure, respectively) in the same search areas.
It can be seen that, in our dataset, the pulmonary veins with higher and lower $RGM_{NAUC}$ were the LSPV and LIPV, respectively. The two halves of the LA, analysed following the \emph{joint-encirclement} approach, exhibit very similar measurements.
Differences across distributions were tested using two-sample t-test: %the null hypothesis was that the distribution of $RGM_{NAUC}$ values corresponding to each vein is the same (one sample set is one vein and the other sample set are the remaining veins). 
the null hypothesis was that $RGM_{NAUC}$ distribution was the same for a given PV comparing to the remaining ones. Statistically significant differences were found for the LIPV ($p = 0.0052$) and the LSPV ($p = 0.029$). 
On the other hand, there were not statistically significant differences between right or left PVs when using the \emph{joint-encirclement} approach ($p = 0.80$).  
Regarding the quantity of gaps around each PV, the LSPV and the LIPV were the veins with the highest and lowest occurrence of gaps, respectively. 
Likewise in the case of the $RGM_{NAUC}$ measurement, a t-test analysis revealed significant differences on the number of gaps with regard to the same veins (LIPV, $p = 0.0059$; LSPV, $p = 0.0092$).
The \emph{joint-encirclement} approach estimated more gaps on the right half of the LA but differences with the left side were not significant ($p=0.27$).

Figure \ref{fig:hists} %shows histograms of $RGM_{NAUC}$, representing the amoung of patients 
the distribution of patients within each $RGM_{NAUC}$ interval.
The first bin in the histogram represents patients with $RGM_{NAUC}$ between 0 and 0.1 (i.e. complete or almost complete encirclement) that would correspond to a more favorable outcome of the PVI procedure. It can be observed that while the LIPV has the highest number of patients with that value of $RGM_{NAUC}$, the LSPV has the lowest.

Results of the regional analysis using the single threshold (T = 3.3 SD) segmentation approach can be seen in Figure \ref{fig:SUM_percentages_colors}. Different 2D SUM plots display the regional distribution of the percentage of patients with at least one gap, the total number of gaps and their length in the whole population.
It can be observed that the region with the highest occurrence of gaps is in between the LSPV and the LAA (LAA ridge).  
 
We also analysed the impact of the chosen segmentation threshold on \emph{RGM} values.
For all thresholds under study, both LIPV and LSPV were significant ($p<0.05$) besides the LSPV when the threshold was 3.3 SD ($p=0.057$), which suggests that, even when the distributions change according to the threshold, the statistical differences remain. 
Figure \ref{fig:boxplots_mean} shows a boxplot representation of the \emph{RGM} distributions corresponding to the multi-threshold approach (left) and to a single-threshold approach with $T = 3.3$ SD.
%Regarding the number of gaps measure, a less stable behaviour was observed: 
The number of gaps had a less stable behaviour than \emph{RGM} values: it was found statistically significant ($p<0.05$) only for some thresholds in the LIPV (3.3 SD, 4 SD) and the LSPV (2 SD, 4 SD, 5 SD). %in LIPV only with thresholds equal to 3.3 SD and 4 SD while in LSPV it was found statistically significant ($p<0.05$) when using thresholds equal to 2 SD, 4 SD and 5 SD. 
In the \emph{joint-encirclement} approach any difference between right and left PVs was found statistically significant in any case.

Finally, the obtained gap quantification indices were compared with the clinical outcome of AF recurrence at follow-up. Recurrence of AF was present in only 11 patients (24\%) who were brought back to the hospital for a redo RFA procedure where reconduction (i.e. electrical conduction recovery after a previous successful isolation) in each PV was assessed. Reconduction was identified in 22 PVs (12\%): 8 cases in RSPV, 5 cases in LSPV, 5 cases in RIPV, and 4 cases in LIPV. As mentioned before, a wide area circumferential ablation (WACA) was performed (i.e. \emph{joint-encirclement} approach) and it is therefore expected to find healthy tissue in the carinas (region between ipsilateral veins) that would increase the \emph{RGM} if the \emph{independent-encirclement} approach is considered. 
However, during the procedure and at the clinicians’ discretion an intervenous line was performed in 18 patients. This was generally performed in the case of more difficult vein isolation in order to achieve the optimal clinical result. 
For this reason, the final gap measure was taken as the minimum among the corresponding \emph{independent-} and \emph{joint-encirclement} approaches (e.g. final-$RGM_{NAUC}$(RSPV) = min($RGM_{NAUC}$(RSPV), $RGM_{NAUC}$(Right PVs))) since we do not know (a priori) which of the two metrics is the smallest one.
Non-significant ($p = 0.523$) slightly higher $RGM_{NAUC}$ values were found among the two groups: mean and SD of 0.258 $\pm$ 0.214 and 0.222 $\pm$ 0.210 for reconnected and not reconnected PVs, respectively. A similar result was obtained when using the single threshold (T = 3.3SD) approach: 0.204 $\pm$ 0.226 vs. 0.156 $\pm$ 0.182, p = 0.328.

All steps in the pipeline were computationally in real-time besides mesh standardization, which required manual interaction, and the non-rigid registration. The execution time of the registration was $18 \pm 7$ min on a desktop computer (Intel i5 3.3 GHz CPU and 16 GB RAM).

\section{Discussion}
\label{sec:discussion}

The purpose of the work presented in this paper was to develop a method to detect and quantify incomplete ablation patterns (gaps) in a reliable, reproducible and observer-independent way. 

Some works can be found in the literature \citep{peters2007detection,badger2010evaluation,halbfass2015lesion} to characterize ablation gaps but all of them rely on visual inspection of complex 3D data by several observers. This approach can be difficult, time-consuming and inaccurate if not performed in a consistent and unbiased way. The proposed method, based on searching minimum distance paths among scar patches, overcomes most of these issues, defining gaps automatically, objectively and with high reproducibility, once a scar pattern in a LA is available. Additionally, the standard parcellation of the LA allows to regionally locate the gaps and study their distribution in different atrial areas.

%Whereas we illustrated our approach with RF-PVI and scar segmentation based on thresholds, 
The method presented in this paper does not rely on any particular ablation technique or scar segmentation approach and it 
%. The proposed \emph{RGM} 
could indeed be used to quantitatively and fairly compare different options for these steps.
Also, we have analysed chronic ablation lesions %(from data acquired 3 months after the procedure) 
but our method could be applied to study acute lesions and their evolution at follow-up. 

The proposed method could help to investigate the relation between electrical and anatomical gaps. However, the suitability of EAM data for gap detection and quantification is limited mainly because of the difficulty of properly correlate them with positions on the LA segmentation acquired from CMR studies. Voltage mapping does not entirely reflect scar formation \citep{kowalski2012histopathologic} for several reasons. For instance, voltage is measured in an area and not at the infinitesimal point where the catheter is touching the wall, and it is highly sensitive to the catheter electrode configuration \citep{josephson2015substrate,blauer2014sensitivity}. The use of EAM to detect scar also requires establishing voltage thresholds for tissue classification, having the same issues as for LGE-MRI thresholds. For all these reasons, direct extrapolation of EAM data to validate LGE-MRI data should be performed with caution, in particular when the two modalities provide contradictory information.

\begin{figure}[t]
\includegraphics[width=\columnwidth]{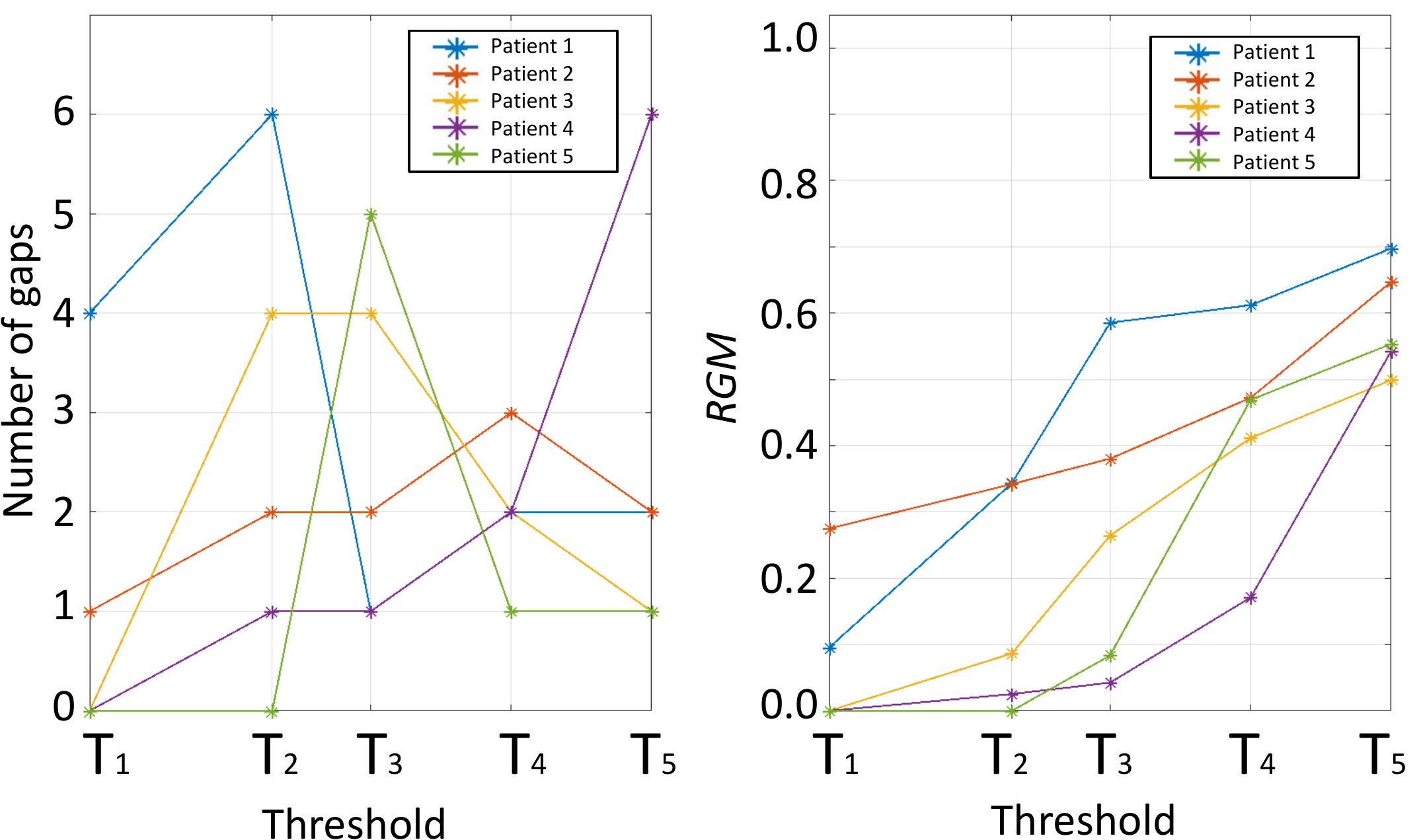} 
\caption{Variation of the number of gaps (left) and \emph{RGM} (right) measurements with regard to the threshold used. Each color corresponds to a different patient. 
}
 \label{fig:evol_thresholds}
\end{figure}

There is high variability in PV morphology (position, orientation, size, thickness) in the whole population. Nonetheless, the proposed \emph{RGM} index is independent of PV variations; it can then be used to compare PV from different patients. We have also found that the \emph{RGM} is more robust to scar segmentation changes than the number of gaps, currently used in gap quantification studies. 
The number of gaps greatly depends on the scar pattern as can be seen in Figure \ref{fig:evol_thresholds} where the evolution of the two measurements with regard to five increasing thresholds is shown. 
Whenever the threshold increases, less tissue is classified as scar and in consequence a higher proportion of gap is expected. The \emph{RGM} (right in Figure \ref{fig:evol_thresholds}) shows this foreseen behavior while the number of gaps measure (left) increases or decreases without any clear tendency in these cases. 

In the studied population, the LIPV and LSPV showed the lowest and highest \emph{RGM} values, respectively.
The higher \emph{RGM} at the LSPV may be partially accounted for by the transverse imaging acquisition plane, with lower resolution between slices. The roof of the LA lies in the transverse plane, and both the imaging and accurate segmentation of these regions, and hence scar detection, are more challenging.
The analysis of the gap probability map (Figure \ref{fig:SUM_percentages_colors}, top) revealed that the most common gap location is the area between the LSPV and the LAA (LAA ridge): almost 70\% of cases had a gap in that region.
In addition to the imaging concerns mentioned above, the elevated number of gaps in that area could also be explained by the presence of a thicker myocardium and the associated difficulty to access with a catheter leading to loss of contact and therefore to non-transmural lesions \citep{galand2016localization,cabrera2009morphological}. The importance of the transmurality of lesions was recently demonstrated by \cite{glover2018preserved}.
Our results are in agreement with the investigation of \cite{furnkranz2010characterization} that found the higher probability of having conduction gaps in the same area (LAA ridge) after cryoballoon ablation.
However, these findings contrast with those of more recent contact force-guided ablation studies with invasive assessment \citep{kautzner2015efficas} where the authors did not find a higher proportion of conduction gaps in that area. 
In contrast to the findings in \cite{halbfass2015lesion} showing that the left PVs had a significantly higher amount of ablation lesions compared with the right PVs (83\% vs 34\%, $p < 0.001$), we did not find significant differences between left and right PVs when considered as ipsilateral pairs. Importantly, the authors used cryoballoon ablation and argued that it is much easier to properly place the cryoballoon into the left PV ostia than into the right ostia which may not be the case in RF ablation.

The absence of a significant relationship in this study between detection of gaps in the CMR-derived ablation lines and recurrence questions the immediate relevance of post-ablation atrial scar imaging. This finding is supported by some recent studies \citep{harrison2015advances,spragg2012initial}, but at odds with others which have demonstrated a significant relationship \citep{badger2010evaluation,bisbal2014cmr,peters2009recurrence,taclas2010relationship}. However, it is important to review carefully those prior publications with positive findings, as they themselves have clearly delineated the limits of the relationship. For example, \cite{peters2009recurrence} found that the degree of scarring around only the RIPV was significant in predicting recurrence, thought to be likely to reflect the technical difficulty in isolating that vein and propensity of triggers to arise from that location. Similarly, \cite{badger2010evaluation} found that only 10 out of 144 ($7\%$) of patients had complete scar encirclement of all PVs, but that there were no recurrences in this group, a statistically significant finding in a small subgroup. The metrics used to assess gaps in this study are arguably more rigorous, but the overall proportion without a gap in both vein pairs’ encirclement is similar ($7\%$ of acquisitions had $>99\%$ PV encirclement of both veins), and in this case was not associated with recurrence.

In this paper we did not compute the RGM for the pre-procedural cases since a comparison between pre- and post-procedural LA scar was already performed in \cite{chubb2018reproducibility} on the same dataset. In pre-procedural cases, a very small proportion of the LA surface was designated as scar and there was a very weak overall correlation between the proportion of pre-ablation scar and the percentage of post-ablation atrial scar ($R^2 = 0.024$, $p=0.02$ across all acquisitions). 
Pre-ablation scar location was therefore interpreted to be unrelated to post-ablation scar location and of minimal significance in further assessment.

The proposed method has a number of limitations mostly related to the pre-processing of the data. First, it highly depends on the LA segmentation step. As mentioned before, we have decided to demonstrate the proposed gap quantification method using a threshold-based approach to detect scar even if thresholding might not be the optimal scar segmentation algorithm as shown in \cite{karim2013evaluation}. The most advanced clinical researchers working on LGE-MRI guided RFA of AF patients are currently using threshold-based methods to segment scar in the LA, as demonstrated in large studies such as the DECAAF clinical trials \citep{marrouche2014association} and in the review paper of \cite{pontecorboli2016use}. As we recently showed \citep{chubb2018reproducibility}, thresholds are likely to differ not only between patients, but also with the time from gadolinium administration. We consider that all these aspects are outside the scope of our work. Therefore, we have used a segmentation algorithm that %was demonstrated to provide accurate segmentations in our database, among those that 
could be easily translated to clinical routine for its daily use. %processing AF patient data. 
The choice was to use the 3.3 SD threshold to detect scar, validated in \cite{harrison2014cardiac} in the first histological study of this type. However, we recognise that their confidence intervals for this threshold were wide. For this reason, we adapted the methodology to simultaneously use a range of thresholds from 2 to 6 SD for the computation of the $RGM_{NAUC}$. 
However, the selection of the most suitable threshold limits and %(i.e., $T_{1}$ and $T_{5}$) as well as 
the number of thresholds considered in between needs to be better investigated.   

Another important limitation of the current implementation of the method is the requirement of a 4 PV configuration in the left atria for its regional parcellation.
This is the most common topology (around 70\% \citep{prasanna2014variations,marom2004variations} but the presence of common trunks or extra PVs is relatively common as well. In our dataset, 90\% of the cases (45 LAs) had 4 PVs while the remaining 10\% (5 LAs) showed some of the topological peculiarities previously mentioned. 

An inaccurate parcellation of the LA could lead to incorrect cutting lines and the impossibility of finding a close encircling path around a PV. In our study, this type of error was identified only in 2 PVs, out of the 180 analysed (from 45 cases).

\section{Conclusion}
\label{sec:conclusion}

In this article, we presented a methodology to detect and quantify incomplete ablation lesions (or gaps) after RF-PVI in a reliable, reproducible and observer-independent way. 
An unambiguous definition of the gap as the minimal portion of healthy tissue around a PV was provided, and showed beneficial specially in complex scenarios where scar lesions are patchy and not continuous. Based on this objective definition of gap, we proposed a quantitative and highly reproducible index, the \emph{RGM}, which represents the proportion of the vein not encircled by scar (i.e. portion of incompleteness). Furthermore, a standard parcellation of the LA was used allowing for regional comparison across patients.  
Additionally, a full characterization of the gaps was provided in terms of their number, position and length.
With the aim of reducing the influence of the scar segmentation process, we proposed a multi-threshold scheme for scar segmentation and the integration of results in the $RGM_{NAUC}$ measure.

The statistical analysis of the results showed that the LIPV and the LSPV had significantly lower and higher $RGM_{NAUC}$ and number of gaps, respectively.
No statistically significant difference was found when considered as ipsilateral (same side) pairs. 
The detailed parcellation of the LA permitted to determine that the LAA ridge was the part of the LA with the highest occurrence of gaps in our population.

We showed the suitability of the proposed method to quantify lesion completeness/incompleteness after PVI and it could therefore be used to compare novel ablation catheters or techniques at full vein isolation efficiency in a consistent and fair way.
Our method could also be used to detect target ablation regions, previously to a redo procedure, favouring its planning and potentially lowering its duration.

\footnotesize \section*{Acknowledgments} This study was partially funded by the Spanish Ministry of Economy and Competitiveness (DPI2015-71640-R) and by the ``Fundaci\'o La Marat\'o de TV3" (n\textsuperscript{o} 20154031). King's College London Medical Engineering Centre is funded by the Wellcome Trust and the Engineering and Physical Sciences Research Council (EPSRC). The research was also supported by the National Institute for Health Research (NIHR) Biomedical Research Centre awards to Guy's and St Thomas' NHS Foundation Trust in partnership with King's College London, by the NIHR Healthcare Technology Co-operative for Cardiovascular Disease at Guy's and St Thomas' NHS Foundation Trust, and by the Cardiovascular HTC. 

\onecolumn
\normalsize
%\section*{\hfil {\color{blue} Appendix. Patient data description?} \hfil}

\twocolumn

\bibliographystyle{elsarticle-harv}
\bibliography{gaps_biblio}

\end{document}